\documentstyle[amsfonts,preprint,aps]{revtex}
\begin{document}

\title{
$J_1-J_2$ quantum Heisenberg antiferromagnet
on the triangular lattice: a group symmetry analysis of order by disorder.
}
\author{
P. Lecheminant \thanks{Laboratoire de Physique Th\'eorique des
    Liquides, Universit\'e P. et M. Curie, boite 121, 4 Place Jussieu, 75252
    Paris Cedex. URA 765 of CNRS}
\thanks{Groupe
    de Physique Statistique, Universit\'e de Cergy-Pontoise,
    95806 Cergy-Pontoise Cedex.},
B. Bernu $ ^* $,
C. Lhuillier $ ^* $, and
L. Pierre\thanks{
    U.F.R. SEGMI, Universit\'e Paris-X, Nanterre, 92001 Nanterre Cedex.}\\
}
\date{\today}
\maketitle

\bibliographystyle{prsty}

\begin{abstract}
On the triangular lattice,
for $J_2/J_1$ between $1/8$ and $1$, the classical Heisenberg model with
first and second neighbor interactions presents four-sublattice
ordered ground-states. Spin-wave calculations of Chubukov and
Jolicoeur\cite{cj92} and Korshunov\cite{k93} suggest that quantum
fluctuations select amongst these states a colinear two-sublattice order.
{}From theoretical requirements, we develop the full symmetry
analysis of the low lying levels of the spin-1/2 Hamiltonian
in the hypotheses of either a four or a two-sublattice order.
We show on the exact spectra of periodic samples ($N=12,16$ and $28$)
how quantum fluctuations select the colinear order from the
four-sublattice order.
\end{abstract}

\newpage
\narrowtext

\section{Introduction}
        Symmetry breaking and the selection of a particular
macroscopic state amongst many degenerate ones result in
part from infinitesimal external causes.
In the case of planar N\'eel order,
the plane of antiferromagnetic ordering, for example, is
chosen by the environment, whereas the possibility
of antiferromagnetic symmetry breaking and the nature of the
antiferromagnetic order are intrinsic and deeply rooted in
the spectral properties of the low lying levels of the
Hamiltonian on a given lattice\cite{a52,blp92,bllp94}. Two features have to
be considered in this respect: ``the ground-state'' and the first
excitations of the system. In the past, interest
has mainly been focussed on the ``first excitations'': the so-called
antiferromagnetic magnons.
The interest in the ground-state has been limited
to the measurement of the order parameter modulus. The approach of this
problem through exact diagonalizations on small samples has led
us to focus on the nature of this ``ground-state'':
the eigenstates of the Heisenberg Hamiltonian on a finite lattice
of $N$ sites are eigenvectors of total spin $S$ and,
in all presently studied cases,
the absolute ground-state is
$S=0$ or $S=1/2$
(depending on the number of sites in the sample). If we
consider the even site samples, the $S=0$ absolute ground-state
is spherically symmetric: it does not break the rotational
symmetry of the Hamiltonian and as such is insufficient to
describe a N\'eel antiferromagnetic state. As underlined
by Anderson in 1952\cite{a52}, the N\'eel symmetry breaking state
arises from a linear combination of a macroscopic number of levels
$\{\tilde E\}$
with different $S$ values which in the thermodynamic limit
collapse to the absolute ground-state faster than the softest magnons.

This set of levels $\{\tilde E\}$
- called QDJS for Quasi Degenerate Joint States in \cite{blp92,bllp94} -
has specific symmetry and dynamical properties which embody the
characteristics of the symmetry breaking phase.
Let us recall that for a finite solid, the low lying levels are
eigenstates of the total momentum and they indeed collapse to the ground
state in the thermodynamic limit faster than the softest phonons. This
is a wave packet of these eigenstates that localizes the center of mass.
Here, in an ordered antiferromagnet, the multiplicity $\{\tilde E\}$ is
associated with the dynamics of the order parameter. In other words, the
knowledge of the symmetry and dynamical properties of this set of
eigenstates yields the nature of the ordered phase.
For the case of a 2d-N\'eel phase, this set of levels
$\{\tilde E\}$ collapses to
the ground-states as $N^{-1}$, that is, faster than other quantities,
in particular faster than the softest magnons which converge to the
ground-state as $N^{-1/2}$\cite{adm93}.
Understanding the symmetry and dynamical properties of these low lying
levels of the Heisenberg Hamiltonian on the triangular lattice leads to
a consistent picture of an ordered ground-state with three sublattice N\'eel
order; this reconciles spin-wave theories and exact diagonalizations
approaches\cite{blp92,bllp94}.

More subtle symmetry breakings still exist
when two or more different kinds of order are classically degenerate. In
the pure classical case, Villain et al\cite{vbcc80} have shown that
thermal fluctuations could select a specific order.
The selected order has softer excitation modes and therefore,
for a given low energy, a larger density of excitations and a
larger entropy: Villain et al\cite{vbcc80} called this mechanism
``order by disorder''. This concept has been rather fruitful
for the studies of classical and quantum antiferromagnets\cite{s82,h89}.

The existence of competing interactions is indeed the main
cause of classical ground-states degeneracy. As a generic example,
one can consider the so-called $J_1-J_2$ model on a triangular
lattice with two competing antiferromagnetic interactions.
This Hamiltonian reads:
\begin{equation}
{\cal H} = 2 J_1 \sum_{<i,j>} {\bf s}_i.{\bf s}_j +
2 J_2 \sum_{<<i,k>>} {\bf s}_i.{\bf s}_k
\label{eq-Heij2}
\end{equation}
where $J_1$ and $J_2 = \alpha J_1$ are positive and the first
and second sums run on the first and second
neighbors, respectively. The classical study of this model has been
developed by Jolicoeur et al\cite{jdgb90}. They have
shown that for small $\alpha$ ($\alpha < 1/8$) the ground-state
corresponds to a three-sublattice N\'eel order with
magnetizations at $120^o$ from each other, whereas for $1/8 < \alpha < 1$,
there is a degeneracy between
a two-sublattice N\'eel and
a four-sublattice N\'eel order (see Fig.\ref{fig-1}).
Chubukov and Jolicoeur\cite{cj92} and Korshunov\cite{k93} have then shown
that quantum fluctuations (evaluated in a spin-wave approach)
 could, like thermal ones,
lift this degeneracy of the classical ground-states and lead to
a selection of the colinear state (see Fig.\ref{fig-1} ).
As usual for spin-1/2 systems, the validity of the spin-wave theory has
to be checked.
The first study of the exact spectrum of Eq.\ref{eq-Heij2} done by
Jolicoeur et al\cite{jdgb90} was not incompatible with this conclusion,
but was
insufficient to yield it immediately. Deutscher and Everts\cite{de93}
found good agreement between spin-wave results
for the colinear state and exact diagonalizations but their sample
geometries were too restricted to fully accommodate the four-sublattice
order. We show in this paper that a study of the complete
dynamical ``ground-state multiplicity'' leads to this conclusion.

In order to understand the origin of this thermodynamical multiplicity
we first
study exactly solvable models which display either four-sublattice order
or colinear order (section II).
Then, on exact spectra of small samples, we show how quantum fluctuations
of increasing wavelength select the colinear order (section III).

\section{Exact solvable quantum models of ordered systems}

These model Hamiltonians are obtained by retaining the
Fourier components of the Heisenberg Hamiltonian which are compatible
either with the two or the four-sublattice order.
In Fourier components,
the Heisenberg Hamiltonian (Eq.\ref{eq-Heij2}) reads:
\begin{equation}
{\cal H} = 6 J_1 \sum_{\bf k} {\bf S}_{\bf k} . {\bf S}_{-\bf k}
\left[\gamma_{\bf k} + \frac{\alpha}{3}
(\cos {\bf k}.(2 {\bf u}_1 + {\bf u}_2) +
\cos {\bf k}.({\bf u}_1 + 2 {\bf u}_2) +
\cos {\bf k}.({\bf u}_2 - {\bf u}_1) ) \right],
\label{eq-h1j2}
\end{equation}
where ${\bf S}_{\bf k} = \frac{1}{\sqrt N} \sum_i {\bf s}_i
\exp i {\bf k} . {\bf r}_i$ and
$\gamma_{\bf k} = 1/3 \sum_{\mu} \cos{\bf k}.{\bf u}_{\mu}$
(${\bf u}_{\mu}$ are three vectors at $120^o$ from each other,
connecting a given site to first neighbors).
In Eq.\ref{eq-h1j2},
the ${\bf k}$-components associated with the  ${\bf k}$-vectors
which keep the sublattices invariant provide
the essential features of the dynamics of the order parameter. We
successively study the case with four-sublattice order and the case with
two-sublattice colinear order.

The four vectors which keep the four-sublattice order invariant are
${\bf k} = {\bf 0}$ and the three middles of the Brillouin
zone boundaries (called in the following ${\bf k}_I$, ${\bf k}_H$
and ${\bf k}_G$).
In this study, we will exclusively
consider finite samples with $N=4p$ sites and with periodic
conditions: these samples do not frustrate the four (nor the two)
sublattice
order and they effectively present the above-mentioned ${\bf k}$
vectors in their Brillouin zone.
It is straightforward
to write the contribution of these Fourier components to ${\cal H}$ in
the form:
\begin{equation}
 ^4{\cal H}_0 = \frac{8}{N}(J_1 + J_2 ) \left( {\bf S}^2
- {\bf S}_A^2 - {\bf S}_B^2 - {\bf S}_C^2 - {\bf S}_D^2 \right),
\label{eq-h0j2}
\end{equation}
where ${\bf S}$ is the total spin operator and the ${\bf S}_{\alpha}$ are
the total spin operator of each sublattice. $^4{\cal H}_0,{\bf S}^2,
S_z,{\bf S}_A^2,{\bf S}_B^2,{\bf S}_C^2$ and ${\bf S}_D^2$
form a set of commuting observables. The eigenstates of $^4{\cal H}_0$
have the following energies:
\begin{eqnarray}
	\nonumber
	^4E(S, S_A, S_B, S_C, S_D) =
		&\frac{8}{N} (J_1 + J_2 ) \left[S(S+1) -  S_A(S_A + 1 ) \right. \\
		& \left. - S_B(S_B + 1 ) - S_C(S_C + 1 ) -S_D(S_D + 1 ) \right]
\label{eq-h01j2}
\end{eqnarray}
where the quantum numbers $S_A,S_B,S_C,S_D$ run from
$0$ to $N/8$ and the total spin results from a coupling of the
four spins $S_A,S_B,S_C,S_D$.

The low lying levels
of Eq.\ref{eq-h01j2} are obtained for $S_A=S_B=S_C=S_D=N/8$:
\begin{equation}
 ^4E_0(S) = -\frac{J_1+J_2}{2}( N+8 ) + \frac{8}{N}(J_1+J_2)S(S+1).
\label{eq-enh02j2}
\end{equation}
These states, which have  maximal sublattice magnetizations
$S_A^2 = S_B^2 = S_C^2 = S_D^2 = (N/8+1)N/8$,
are the {\it rotationally invariant projections of the bare
N\'eel states with four sublattices}.
This is the single physical origin of all properties of $\{^4\tilde E\}$.
These levels have an energy
collapsing to the absolute ground-state as $N^{-1}$ justifying the name
of tower of states or
``ground-state multiplicity'' given to $\{^4\tilde E\}$.
In this exactly solvable model there are no quantum fluctuations
to renormalize the sublattice magnetization;
quantum fluctuations will be introduced by the discarded part of the
Hamiltonian (Eq.\ref{eq-h1j2}).

As we will now show, this multiplicity $\{^4\tilde E\}$ can be
entirely and  uniquely described by its
symmetry properties  under spin rotation and transformation of
the space group of the lattice.

Let us begin by the $SU(2)$ properties induced by the fact that
these states represent the coupling of four identical spins.
 The degeneracy of each $S$ level is $(2S+1)N_S$ where
the factor $(2S+1)$ comes from the magnetic degeneracy and
$N_S$ is the number of different couplings of four spins, each of length $N/8$,
giving a total spin $S$. This number is readily evaluated
by using the decomposition of the product of four spins $n/8$
representations of $SU(2)$ (${\cal D}^{N/8}$):
\begin{equation}
\{^4\tilde E\}= {\cal D}^{N/8}\otimes  {\cal D}^{N/8}\otimes  {\cal
D}^{N/8}\otimes  {\cal D}^{N/8}
\label{e4}
\end{equation}
in  spin $S$ irreducible representations
(${\cal D}^S$).
One obtains:
\begin{equation}
\left\{
\begin{array}{lll}
\displaystyle N_S &= & \displaystyle \frac{1}{2}\left(-3 S^2 + S(N+1)
+2 +\frac{N}{2}
			  \right)~~~~ {\rm for~~~} S\le \frac{N}{4}, \\
\\
\displaystyle      & = & \displaystyle \frac{1}{2}
\left(\frac{N}{2}-S+1\right) \left(\frac{N}{2}-S+2\right)
				~~~~ {\rm for~~~} S\ge \frac{N}{4}+1.
\end{array}
\right.
\label{eq-mulsj2}
\end{equation}
Note that this degeneracy depends both
on $S$ and $N$ and not only on the total spin $S$ as
is the case for two or three-sublattice problems. In fact,
in the latter two cases, which describe N\'eel
order on a square or triangular lattice, the objects to
be considered stem from the coupling of two or three
angular momenta: they have perfect counterparts in
the orbital three dimensional world which are rigid rotators and
tops with well known quantum numbers, depending only on $S$.
More generally, a N\'eel order with $p$ sublattices on a finite sample
of $N$ spins gives rise to a ``ground-state multiplicity'' of the order
of $N^p$.

The determination of the space symmetries of these eigenstates
allows a complete specification of $\{ ^4\tilde E\}$.
The four-sublattice order is invariant in a two-fold rotation
($\cal{R_{\pi}}$): thus the eigenstates of  $\{ ^4\tilde E\}$ belong to
the trivial representation of $C_2$.
As it arises from the coupling of four identical spins,
this subset of levels forms a representation space of $S_4$,
the permutation group of four elements. The eigenstates of $\{ ^4\tilde E\}$
could thus be labeled by the irreducible representations of  $S_4$
(see Table \ref{table-1}). Indeed, the complete analysis of all
the eigenstates of Eq.\ref{eq-Heij2} is usually
done through the more general point
of view of the space group of the lattice. But it is
straightforward to show that in the four-sublattice subset of
solutions, each element of the space group maps onto a permutation of $S_4$:
one step translations map onto products of transpositions as $(A,B)(C,D)$,
three-fold rotations onto circular permutations of three
sublattices $(A,B,C)$ and so on. The
complete mapping of the space symmetries of the four-sublattice
order onto the permutations of $S_4$ is given in
Table \ref{table-1} together with the character table of $S_4$. Each
irreducible representation of $S_4$ can thus be characterized
in terms of its space symmetry properties. As noted above they are all
invariant in $\cal{R_{\pi}}$. Analysis of the properties under
translation shows that $\Gamma_1,\Gamma_2$ and $\Gamma_3$ correspond
to the wave-vector ${\bf 0}$, whereas $\Gamma_4$ and
$\Gamma_5$ have a wave-vector
${\bf k}_H, {\bf k}_I$ or ${\bf k}_J$.
$\Gamma_1$ and $\Gamma_2$ are invariant under $C_3$, whereas $\Gamma_3$
is associated with the two complex representations of this
same group. Finally, $\Gamma_1$ and $\Gamma_4$ are
even under axial symmetry whereas $\Gamma_2$ and $\Gamma_5$ are odd.
The number of replicas of $\Gamma_i$ that should appear
for each $S$ is then computed in the $S,M_S$ subspace
with the help of the trace of the permutations of $S_4$:
\begin{equation}
	\label{eq-ngammai}
	n_{\Gamma_i}^{(S)} = \frac{1}{24} \sum_k {\rm Tr}(R_k|_S) \chi_i(k) N_{\rm
	el}(k)
\label{noc}
\end{equation}
where $R_k$ is an element of the class $k$ of  $S_4$, $N_{\rm el}(k)$
is the number
of elements in this class and $\chi_i(k)$ is the character of the class $k$
in the irreducible
representation $\Gamma_i$ (see Table \ref{table-1}).
The values of the traces for a given total spin $S$ are then found
as:
\begin{equation}
{\rm Tr} \left( (A,B,C)\bigg |_S \right) =
{\rm Tr} \left( (A,B,C)\bigg |_{M_S=S} \right) -
{\rm Tr} \left( (A,B,C)\bigg |_{M_S=S+1} \right).
\label{eq-tr2sz4}
\end{equation}
In each $M_S$ subspace of $\{ ^4\tilde E\}$,
it is straightforward to find the trace of the elements of  $S_4$:
\begin{equation}
\left\{
\begin{array}{lll}
\displaystyle {\rm Tr} \left(I_d\bigg |_{M_S} \right) &=&
	\displaystyle	\sum_{t,v,x,y=-N/8}^{N/8} \delta_{t+v+x+y ,M_S} \\
\displaystyle {\rm Tr} \left((A,B) (C,D)\bigg|_{M_S} \right) &=&
	\displaystyle	\sum_{t,v=-N/8}^{N/8} \delta_{2t+2v, M_S}\\
\displaystyle {\rm Tr} \left((A,B,C)\bigg |_{M_S} \right) &=&
	\displaystyle	\sum_{t,v=-N/8}^{N/8} \delta_{3t+v,M_S} \\
\displaystyle {\rm Tr} \left((A,B)\bigg |_{M_S} \right) &=&
	\displaystyle	\sum_{t,v,x=-N/8}^{N/8} \delta_{2t+v+x,M_S} \\
\displaystyle {\rm Tr} \left((A,B,C,D)\bigg |_{M_S} \right) &=&
	\displaystyle	\sum_{t=-N/8}^{N/8} \delta_{4t,M_S}
\end{array}
\right.
\label{eq-holpri}
\end{equation}
 where $t,v,x,y$ are the $z$-components of the total spin
of each sublattice (constrained to vary between $N/8$ and $-N/8$) and
$\delta_{i,j}$ denotes the Kronecker symbol.
Using Eqs.\ref{noc},\ref{eq-tr2sz4},\ref{eq-holpri}
one readily obtains the number of occurrences of each $\Gamma_i$
for each  $S$ subset of $\{ ^4\tilde E\}$ (Table \ref{table-2}).
We have thus obtained
the complete determination (all quantum numbers, and all the
degeneracies) of the family of low lying levels describing the
ground-state multiplicity of the four-sublattice N\'eel solutions.

Let us now consider the  colinear solutions (Fig.\ref{fig-1}).
They are particular
solutions of the four-sublattice case and  we will rapidly go through
the same scheme of analysis, indicating mainly the new points.
The two vectors which keep the two sublattices invariant are
${\bf 0}$ and the middle of one side of the Brillouin
zone (the vectors ${\bf k}_I$, ${\bf k}_H$ and ${\bf k}_G$
correspond respectively to the
colinear solutions $(a)$, $(b)$ and $(c)$ in Fig.\ref{fig-1}).
Extracting  a specific set of two wave-vectors
from Eq.\ref{eq-h1j2}, we find the following
contribution to the total Hamiltonian:
\begin{equation}
 ^2{\cal H}_0 = \frac{8}{N}(J_1 + J_2) \left[ {\bf S}^2
- \frac{1}{2} ( {\bf S}_\alpha^2 + {\bf S}_\beta^2 ) \right].
\label{eq-colj2}
\end{equation}
The corresponding low energy spectrum for $S_{\alpha}=S_{\beta}=N/4$ is:
\begin{equation}
 ^2E_0(S) = -\frac{J_1+J_2}{2}( N+8 ) + \frac{8}{N}(J_1+J_2)S(S+1)
\label{eq-col1j2}
\end{equation}
and is degenerate with the four-sublattice low energy spectrum
(see Eq.\ref{eq-enh02j2}).
Here, the two sublattices have maximal spins
$S_{\alpha }=S_{\beta}=N/4$. These new solutions
arise from the
symmetric coupling of the spins of two sublattices of the
four-sublattice order: $ \bf{S}_{\alpha }= \bf{S}_A+ \bf{S}_B$ or
 $\bf{S}_{\alpha }= \bf{S}_A+ \bf{S}_C$ or
 $\bf{S}_{\alpha}= \bf{S}_A+ \bf{S}_D$
with the counterparts for $ \bf{S}_{\beta}$.
As there are three ways to do this coupling,
the colinear solutions have a $Z_3$ degeneracy.
The representation space is thus the sum of three products
${\cal D}^{N/4}\otimes{\cal D}^{N/4}$. It is not a direct
sum since ${\cal D}^{N/4}(A,B) \otimes {\cal D}^{N/4}(C,D)$ and
${\cal D}^{N/4}(A,C) \otimes {\cal D}^{N/4}(B,D)$ have in
common the same (symmetric) irreducible representation
with a total spin $N/2$. On a $N$-sample, the representation
space of the ground-state of the colinear solution is:
\begin{equation}
\{^2\tilde E\} = 3 {\cal D}^{S=0} \oplus 3 {\cal D}^{S=1} \oplus ....
\oplus 3{\cal D}^{S=N/2-1} \oplus {\cal D}^{S=N/2}.
\label{eq-spacetwo}
\end{equation}
The degeneracy is thus $3(2S+1)$ for all $S$ values except for $S=N/2$
where it is $(2S+1)$.

As for the four-sublattice order, the space-group analysis is done as
for the
two-sublattice order, but the number of occurrences of each irreducible
representations  $\Gamma_i$ is now different since the space
$\{ ^2\tilde E\}$ is smaller than $\{ ^4\tilde E\}$. The calculation
could be done along the same lines as for the four-sublattice order.
The problem, however, is much simpler because for each $S$ value
there are only
three replicas of ${\cal D}^S$ arising from the $Z_3$ group
(Eq.\ref{eq-spacetwo} and Fig.\ref{fig-1}). This allows direct
computation of the permutation traces in each $S$
subset of $\{ ^2\tilde E\}$.
Using the coupling rules of two angular momenta
(and in particular the fact that the $S$ eigenstate resulting
from the coupling of two integer spins changes sign as
$(-1)^S$ with the interchange of the two parent spins) one
obtains (for $S \neq N/2$):
\begin{equation}
\left\{
	\begin{array}{lll}
		\displaystyle {\rm Tr} \left( I_d \bigg |_S\right) &=& 3\\
		\displaystyle {\rm Tr} \left( (A,B)(C,D) \bigg |_S\right)
																			&=& 1 + 2(-1)^S\\
		\displaystyle {\rm Tr} \left( (A,B,C)\bigg |_S\right) &=& 0\\
		\displaystyle {\rm Tr} \left( (A,B)\bigg |_S\right) &=& 1\\
		\displaystyle {\rm Tr} \left( (A,B,C,D)\bigg |_S\right) &=& (-1)^S
	\end{array}
\right.
\label{eq-trij24}
\end{equation}
Therefore, the colinear solution is simply
characterized by $\Gamma_1$ and $\Gamma_3$
for even $S$ and $\Gamma_4$ for odd $S$.

{}From these equations
(Eqs.\ref{noc},\ref{eq-tr2sz4},\ref{eq-holpri},\ref{eq-trij24}),
the symmetries of all states of
the tower are fully
determined both for the four-sublattice order $\{ ^4\tilde E\}$
and for the colinear order $\{ ^2\tilde E\}$.

Going back now to the original $(J_1-J_2)$ model, we have to account for
quantum fluctuations generated by the discarded part of $\cal H$. This
perturbation does not commute with sublattice total spins and
consequently reduces the sublattice magnetization. Nevertheless, it
preserves all the symmetries of the N\'eel state and thus also the ones of
the levels of $\{ ^4\tilde E\}$ or $\{ ^2\tilde E\}$.
Then the question is : do quantum fluctuations conserve qualitatively
the dynamics of these levels or not ? If these levels remain the low lying
ones of the exact spectra with overall dynamics qualitatively similar
to that of
the bare N\'eel state (Eqs. \ref{eq-enh02j2},\ref{eq-col1j2}),
then the quantum model will be ordered at $T=0$.
By qualitatively,
we mean that the leading term of the energy
of the exact subset $\{E\}$ behaves as
$\beta \frac{8}{N} (J_1+J_2) S(S+1)$, where $\beta$ is a
renormalization factor.  This factor is related to the spherical homogeneous
susceptibility of the sample \cite{bllp94},
even if, in general, the tensor of susceptibilities is not spherical because
quantum fluctuations lift the degeneracies of $\{ ^4\tilde E\}$ of
the exactly solvable model.

\section{Exact spectra of small periodic samples}

We have determined the low (and high) energy levels of the $J_1-J_2$
Hamiltonian
in each irreducible representation of $SU(2)$ and of the space group of
the triangular lattice for small periodic samples with $N=12,16$ and $28$.
The spectra are displayed in Fig.\ref{fig-N=16} and Fig.\ref{fig-N=28}.
We directly see in the upper parts of these figures the set
of QDJS (``ground-state multiplicity'') well separated from the set of levels
corresponding to the one magnon excitations. We have verified that
the QDJS form  a set of levels with the exact properties of the above defined
$\{ ^4\tilde E\}$ subset.
The action of quantum fluctuations could then be read in
the  lower parts of the figures. As expected,
quantum fluctuations lift the degeneracies
which are present in the exactly solvable model
and stabilize the eigenstates with the lower $S$ values.
Nevertheless, the low lying
energies per site still group around a line of equation
$E_\infty+8\beta S(S+1)(J_1+J_2)/N^2$ with $\beta =1.004$ (resp. $1.055$)
for $N=16$ (resp. $28$).
The  number and space symmetries of these levels
for each $S$ and $N$ value are exactly those required by
the above analysis of the four-sublattice N\'eel order.
Moreover, it is already visible on the $N=16$ sample and quite clear
on the $N=28$ sample that a dichotomy
appears in this family (see Fig.\ref{fig-QDJ}).
The lowest levels of this tower
of states appear to be $\Gamma_1,\Gamma_3$ or $\Gamma_4$
representations depending on the parity of the total
spin. They precisely build the family  $\{ ^2\tilde E\}$ of isotropic
projections of the colinear solutions given above (Eq.\ref{eq-trij24}).

We see in Fig.\ref{fig-QDJ} that the difference between the energy
per bond of the colinear states and that of the other states of
the four-sublattice order roughly increases by a factor 4 from the
$N=16$ to the $N=28$ sample. This strongly suggests that
the four-sublattice order will disappear in the thermodynamic limit
and only the colinear order will persist.
This result supports the conclusion of the
spin-wave expansion\cite{jdgb90,cj92,k93} concerning
the selection of the colinear state in the $J_1-J_2$ model for
$1/8 < \alpha < 1$ for spins $1/2$.

\section{Concluding remarks}

It appears from the
two situations that we have studied (triangular
Heisenberg model\cite{blp92,bllp94} and this model) that
the symmetry and dynamical analysis of the low lying levels
of a Hamiltonian likely to exhibit ordered solutions
give rather straightforward answer to the kind
of order to be expected. The method is rapid, powerful
and unbiased: it does not require any {\it a priori} symmetry breaking
choice: if a specific order is selected, one
should see it directly on the exact spectra. Moreover, as
it is essentially exact, there are no questions relative
to the convergence of the expansion as in the spin-wave
approach. On the other hand, as the sizes amenable to
computation are limited, there is, in the exact
approach, a cut-off of the long wavelength fluctuations.
Results so obtained should thus be examined in
light of a finite size scaling analysis.
The present work nevertheless shows that it is not necessary to invoke
quantum fluctuations with very long wavelengths to select the colinear order.

{\bf Acknowledgements}:

We have benefited from a grant of computer time at Centre de Calcul
Vectoriel pour la Recherche (CCVR), Palaiseau, France.
We would like to thank P. Von Tassel for a careful
reading of the manuscript.
Data of the spectrum are available under request at bernu@lptl.jussieu.fr

\newpage

\begin{table}
\begin{center}
\begin{tabular}{|c|c c c c c |}
\hline
$S_4$ & $ I $ & $(A,B)(C,D)$ & $(A,B,C)$ & $(A,B)$ & $(A,B,C,D)$\\
${\cal G}$ & $I$ & $t$ & ${\cal R}_{2\pi/3}$ & $\sigma$
& ${\cal R}_{2\pi/3}^{'} \sigma$ \\
  $ N_{el}$ & 1 & 3 & 8 & 6 & 6\\
  \hline
   $\Gamma_1 $ & $ 1 $ & $ 1 $ & $ 1 $ & $ 1 $ & $ 1 $ \\
   $\Gamma_2 $ & $ 1 $ & $ 1 $ & $ 1 $ & $-1 $ & $-1 $ \\
   $\Gamma_3 $ & $ 2 $ & $ 2 $ & $-1 $ & $ 0 $ & $ 0 $ \\
   $\Gamma_4 $ & $ 3 $ & $-1 $ & $ 0 $ & $ 1 $ & $-1 $ \\
   $\Gamma_5 $ & $ 3 $ & $-1 $ & $ 0 $ & $-1 $ & $ 1 $ \\
\hline
\end{tabular}
\end{center}
\caption[99]{Character table of the permutation group $S_4$.
First line indicates classes of permutations.
Second line gives an element of the space symmetry class
corresponding to the class of permutation. These space symmetries are: $t$
the one step translation  ($A\to B$),  ${\cal R}_{2\pi/3}$ (resp.
${\cal R}_{2\pi/3}^{'}$)
the three-fold rotation around a site of the
$D$ (resp. $B$)-sublattice, and $\sigma$ the axial symmetry
keeping invariant $C$ and $D$.
$N_{el}$ is the number of elements in each class.}
\label{table-1}
\end{table}

\begin{table}
	\begin{center}
		\begin{tabular}{|c|c c c c c c c c c|}
			\hline
			$N=16$ &&&&&&&&& \\
			$S$ & 0 & 1 & 2 & 3 & 4 & 5 & 6 & 7 & 8 \\
			\hline
			$n_{\Gamma_1}(S)$ &1 & 0 & 2 & 0 & 2 & 1 & 1 & 0 & 1 \\
			$n_{\Gamma_2}(S)$ &0 & 0 & 1 & 0 & 0 & 0 & 0 & 0 & 0 \\
			$n_{\Gamma_3}(S)$ &2 & 0 & 2 & 1 & 2 & 0 & 1 & 0 & 0 \\
			$n_{\Gamma_4}(S)$ &0 & 2 & 2 & 3 & 2 & 2 & 1 & 1 & 0 \\
			$n_{\Gamma_5}(S)$ &0 & 2 & 1 & 2 & 1 & 1 & 0 & 0 & 0 \\
			\hline
		\end{tabular}
		\begin{tabular}{|c|c c c c c c c c c c c c c c c|}
			\hline
			$N=28$ &&&&&&&&&&&&&&& \\
			$S$ & 0 & 1 & 2 & 3 & 4 & 5 & 6 & 7 & 8 & 9 & 10 & 11 & 12 & 13
			& 14\\
			\hline
			$n_{\Gamma_1}(S)+n_{\Gamma_2}(S)$& 2&0&5&1&5&3&4&2&4&1&2&1&1&0&1\\
			$n_{\Gamma_3}(S)$& 3&0&4&2&5&2&5&2&3&1&2&0&1&0&0\\
			$n_{\Gamma_4}(S)+n_{\Gamma_5}(S)$& 0&7&6&11&9&12&9&10&6&6&3&3&1&1&0\\
			\hline
      \end{tabular}
	\end{center}
\caption[99]{Number of occurrences $n_{\Gamma_i}(S)$ of each irreducible
representation $\Gamma_i$ with respect to the total spin $S$.
For $N=28$,
$n_{\Gamma_1}$ and $n_{\Gamma_2}$ as well as
$n_{\Gamma_4}$ and $n_{\Gamma_5}$ have been added because this sample
does not present any axial symmetry.}
\label{table-2}
\end{table}

\begin{figure}
\caption[99]{
Top: four-sublattice classical ground-state.
Spins in the sublattices $A$ and $B$,
as well as spins in $C$ and $D$,
make an angle $2\theta$.
The plane of the spins of $A$ and $B$ makes an angle $\phi$ with the
plane of the spins of $C$ and $D$.
Bottom:  the colinear solutions with the three possible arrangements
(in this case, classical spins in sublattices A and B are antiparallel).}
\label{fig-1}
\end{figure}

\begin{figure}
\caption[99]{
Top: complete spectrum of the $N=16$ periodic sample with respect to
${\bf S}^2$ for $J_2/J_1=0.7$.
Bottom: enlargement of the difference between the exact spectrum and
the energy of the low lying levels of the model Hamiltonian
(Eq.\ref{eq-enh02j2} or Eq.\ref{eq-col1j2}).
The  ground-state multiplicity  $\{ ^4\tilde E\}$
is well separated from the magnons.}
\label{fig-N=16}
\end{figure}

\begin{figure}
\caption[99]{
Partial spectrum of the $N=28$ periodic sample
(same legend as for Fig.\ref{fig-N=16}). Bottom:
the tower of states of the four-sublattice order
$\{ ^4\tilde E\}$ lays under the dashed line.
Above appear the first magnons. Above the dotted line
are represented the first excited homogeneous states.
In the magnon multiplicity (${\bf k} \neq {\bf 0},
{\bf k}_H,{\bf k}_I$ or ${\bf k}_J$), for $S\le 5$,
only the 5 lowest states of each irreducible representation
have been computed.}
\label{fig-N=28}
\end{figure}

\begin{figure}
\caption[99]{
Enlargement of the $N=16$ and $N=28$ QDJS. A global contribution
$ \beta E_0(S)$ is subtracted from the exact spectrum. This
contribution describes the overall dynamics of the order parameter
in this finite sample, $\beta $ measures the renormalization of
this dynamics by quantum fluctuations (see text and Ref.\cite{bllp94}).
The bars represent
eigenstates which belong both to $\{ ^2\tilde E\}$ and $\{ ^4\tilde E\}$.
The triangles indicate states which belong to $\{ ^4\tilde E\}$  but
not to $\{ ^2\tilde E\}$. We see that, with increasing sizes, the
tower of states of the colinear order separates
from the four-sublattice order. For $N=28$,
the  two  states of $\{ ^2\tilde E\}$ with even $S$ are quasi degenerate and
cannot be distinguished at the scale of the figure.}
\label{fig-QDJ}
\end{figure}

\end{document}